\pdfoutput=1
\documentclass[prl,aps,twocolumn,showpacs,superscriptaddress]{revtex4}
\usepackage{graphicx,amssymb,amsmath,color,psfrag}
\usepackage{amsmath}
\usepackage{amsthm}
\usepackage{amsfonts}
\usepackage{algorithmic}
\usepackage{enumerate}
\usepackage{latexsym}

\DeclareMathOperator{\sinc}{sinc}
\input{tcilatex}
\begin{document}

\title{Observation of Intensity-Intensity Correlation Speckle Patterns with
Thermal Light}
\author{Li-Gang Wang}
\affiliation{Institute for Quantum Science and Engineering (IQSE) and Department of
Physics and Astronomy, Texas A$\&$M University, College Station, Texas
77843-4242, USA}
\affiliation{Department of Physics, Zhejiang University, Hangzhou 310027, China}
\affiliation{The National Center for Applied Physics, KACST, P. O. Box 6086, Riyadh
11442, Saudi Arabia}
\author{M. Al-Amri}
\affiliation{Institute for Quantum Science and Engineering (IQSE) and Department of
Physics and Astronomy, Texas A$\&$M University, College Station, Texas
77843-4242, USA}
\affiliation{The National Center for Applied Physics, KACST, P. O. Box 6086, Riyadh
11442, Saudi Arabia}
\affiliation{Beijing Computational Science Research Center, Beijing, 100084, China}
\author{M. Suhail Zubairy}
\affiliation{Institute for Quantum Science and Engineering (IQSE) and Department of
Physics and Astronomy, Texas A$\&$M University, College Station, Texas
77843-4242, USA}
\affiliation{The National Center for Applied Physics, KACST, P. O. Box 6086, Riyadh
11442, Saudi Arabia}
\affiliation{Beijing Computational Science Research Center, Beijing, 100084, China}

\begin{abstract}
In traditional Hanbury Brown and Twiss (HBT) schemes, the thermal
intensity-intensity correlations are phase insensitive. Here we propose a
modified HBT scheme with phase conjugation to demonstrate the
phase-sensitive and nonfactorizable features for thermal intensity-intensity
correlation speckle. Our scheme leads to results that are similar to those
of the two-photon speckle. We discuss the possibility of the experimental
realization. The results provide us a deeper insight of the thermal
correlations and may lead to more significant applications in imaging and
speckle technologies.




\end{abstract}

\date{\today }
\maketitle

Optical speckle usually refers to the random interference phenomenon that
happens when coherent light fields are reflected from (or pass through) a
disorder scattering medium \cite{Goodman2006}. This phenomenon has been
recognized to be the manifestations of the random\ characteristics (e. g.,
randomly varying phase and amplitude) of a scattering medium. Various
applications have been developed to make use of the speckle phenomena in
fields ranging from astronomy to random lasers \cite%
{Goodman2006,Bortolozzo2011}.

To observe optical speckle, one often needs the light source with good
spatial coherence. It is widely believed that there is no speckle effect for
thermal or incoherent light fields. The conventional speckle is usually
described by the scattered intensity, which is regarded as the one-photon
probability density. Recently, the concept of two-photon speckle, described
via a two-photon probability density, was developed elegantly within the
theory of quantum correlations \cite{Beenakker2009,Cande2013} and was
demonstrated experimentally via the coincidence measurements (or
intensity-intensity correlation measurements) \cite{Peeters2010,
vanExter2012, LorenzoPires2012}. These studies are important to directly
visualize the spatial structure of the entanglement in the scattered light.

Recently, there have also been a series of theoretical and experimental
investigations \cite%
{Bennink2002,Bennink2004,Gatti2004,ZhuSY2004,WangKG2004,Ferri2005,ZhaiYH2005}
with pseudothermal or true thermal light, on ghost imaging, ghost
diffraction and interference due to certain similarity between a two-photon
source and an incoherent light \cite{Saleh2000}. Until now, the
intensity-intensity correlation speckle for thermal and incoherent light has
remained unexplored. It was claimed that the nonfactorizable features in
two-photon speckle are not present for thermal light \cite{Peeters2010},
since thermal correlations are phase insensitive \cite{Erkmen2008}.

In this Letter, we propose a modified Hanbury-Brown and Twiss (HBT) scheme
to change thermal correlations for observing the intensity-intensity
correlation speckle for thermal light. Our scheme, same as two-photon
speckle \cite{Peeters2010,LorenzoPires2012}, is different from those based
on ghost imaging. The thermal photons in our case pass through a common
transmission mask (TM), and the light source here is thermal light not the
entangled two-photon source.

\begin{figure}[b]
\includegraphics[width=5.0cm]{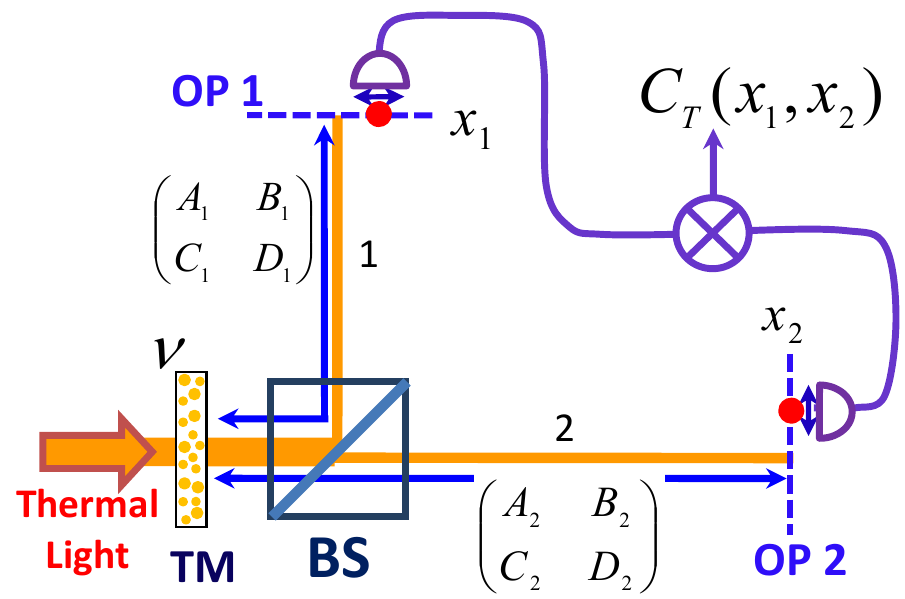}
\caption{(color online). The traditional HBT scheme. The TM is in front of
the beam splitter (BS), and the intensities on the output planes (OPs) 1 and
2 are correlated by a correlator. Optical paths 1 (2) from the TM to the OPs
1 (2) are characterized by the $2\times 2$ ray transfer matrices.}
\label{Fig:fig1}
\end{figure}

We first briefly discuss the traditional HBT scheme \cite%
{HanburyBrown1956,AlQasimi2010}, see Fig. 1. The light passes through the
TM, and then it is divided into two paths by the beam splitter (BS). It is
known that, for thermal or incoherent sources obeying Gaussian statistics,
the intensity-intensity correlation $C_{T}(x_{1},x_{2})$ is expressed by
Siegert relation \cite{Saleh1978}%
\begin{equation}
C_{T}(x_{1},x_{2})=\left\langle I_{T}(x_{1})\right\rangle \left\langle
I_{T}(x_{2})\right\rangle +\left\vert W_{T}(x_{1},x_{2})\right\vert ^{2},
\end{equation}%
where $\left\langle I_{T}(x_{j})\right\rangle $ ($j=1,2$) are the average
intensities on the output planes, $W_{T}(x_{1},x_{2})$ is the cross-spectral
density between the two output planes, and they are respectively given by 
\cite{Saleh2000}%
\begin{gather}
\left\langle I_{T}(x_{j})\right\rangle \!=\!\!\!\iint \!\!\!W_{i}(\nu
_{1},\nu _{2})h_{j}^{\ast }(\nu _{1},x_{j})h_{j}(\nu _{2},x_{j})d\nu
_{1}d\nu _{2},  \label{G0III} \\
\!W_{T}(x\!_{1}\!,x_{2})\!=\!\!\!\iint \!\!\!W_{i}(\nu _{1}\!,\!\nu
_{2})h_{1}^{\ast }(\nu _{1}\!,\!x_{1})h_{2}(\nu _{2}\!,\!x_{2})d\nu _{1}d\nu
_{2}.  \label{G1function}
\end{gather}%
Here $W_{i}(\nu _{1},\nu _{2})\equiv \left\langle E_{i}^{\ast }(\nu
_{1})E_{i}(\nu _{2})\right\rangle $ is the initial cross-spectral density of
the input random light fields $E_{i}(\nu )$ at the TM. The impulse response
functions $h_{j}(\nu ,x_{j})$, from the Collins' formula, can be expressed
as \cite{Collins1970,WangSMZhaoDM} 
\begin{equation}
h_{j}(\nu ,x_{j})=t(\nu )(\frac{-i}{\lambda B_{j}})^{\frac{1}{2}}e^{\frac{%
i\pi }{\lambda B_{j}}(A_{j}\nu ^{2}-2\nu x_{j}+D_{j}x_{j}^{2})}
\label{ah1h2}
\end{equation}%
under the paraxial approximation, where $\lambda $ is the wavelength, $A_{j}$%
, $B_{j}$, and $D_{j}$ are the elements of the $2\times 2$ ray transfer
matrices $\left( 
\begin{array}{cc}
A_{j} & B_{j} \\ 
C_{j} & D_{j}%
\end{array}%
\right) $ describing the linear optical systems \cite{Milonni2010} from the
TM to the output planes, and $t(\nu )$ is the complex transmission
coefficient of the TM.

For simplicity, both optical paths 1 and 2 are assumed to be within the
range of Fraunhofer diffraction \cite{WangSMZhaoDM}, i. e., $A_{j}=0$.
Meanwhile, the input light is a thermal or incoherent source, i. e., $%
W_{i}(u_{1},u_{2})=I_{0}\delta (u_{1}-u_{2})$ with $I_{0}$ a constant.
Therefore, $C_{T}(x_{1},x_{2})$ can be written as 
\begin{equation}
C_{T}(x_{1},x_{2})=\left\langle I_{T}(x_{1})\right\rangle \left\langle
I_{T}(x_{2})\right\rangle \left[ 1+\mu _{T}(x_{1},x_{2})\right] ,
\label{CCCTT}
\end{equation}%
where%
\begin{equation}
\mu _{T}(x_{1},x_{2})=\frac{1}{N_{0}^{2}}\left\vert \tciFourier _{1}\left[
\left\vert t(u)\right\vert ^{2}\right] (\frac{x_{2}}{\lambda B_{2}}-\frac{%
x_{1}}{\lambda B_{1}})\right\vert ^{2}  \label{UUUINC}
\end{equation}%
is the normalized \textit{phase-insensitive} shape function. This shape
function is only related to $\left\vert t(\nu )\right\vert ^{2}$, $%
\left\langle I_{T}(x_{j})\right\rangle =I_{0}N_{0}(\lambda \left\vert
B_{j}\right\vert )^{-1}$ with $N_{0}=\int \left\vert t(\nu )\right\vert
^{2}d\nu $, and $\tciFourier _{1}$ denotes the one-dimensional Fourier
transform of $\left\vert t(\nu )\right\vert ^{2}$ with the argument of $%
\frac{x_{2}}{\lambda B_{2}}-\frac{x_{1}}{\lambda B_{1}}$. It is clear that $%
\mu _{T}(x_{1},x_{2})$ contains only the partial information of $t(\nu )$
[i. e., the amplitude of $t(\nu )$], and it does not have any phase
information of $t(\nu )$. Therefore, the thermal intensity-intensity
correlations based on the traditional HBT scheme are essentially phase
insensitive \cite{Peeters2010,Erkmen2008}. It should be emphasized that both 
$\left\langle I_{T}(x_{1})\right\rangle $ and $\left\langle
I_{T}(x_{2})\right\rangle $ are uniform and have no any information of $%
t(\nu )$ for completely incoherent fields.

\begin{figure}[tbp]
\includegraphics[width=8.5cm]{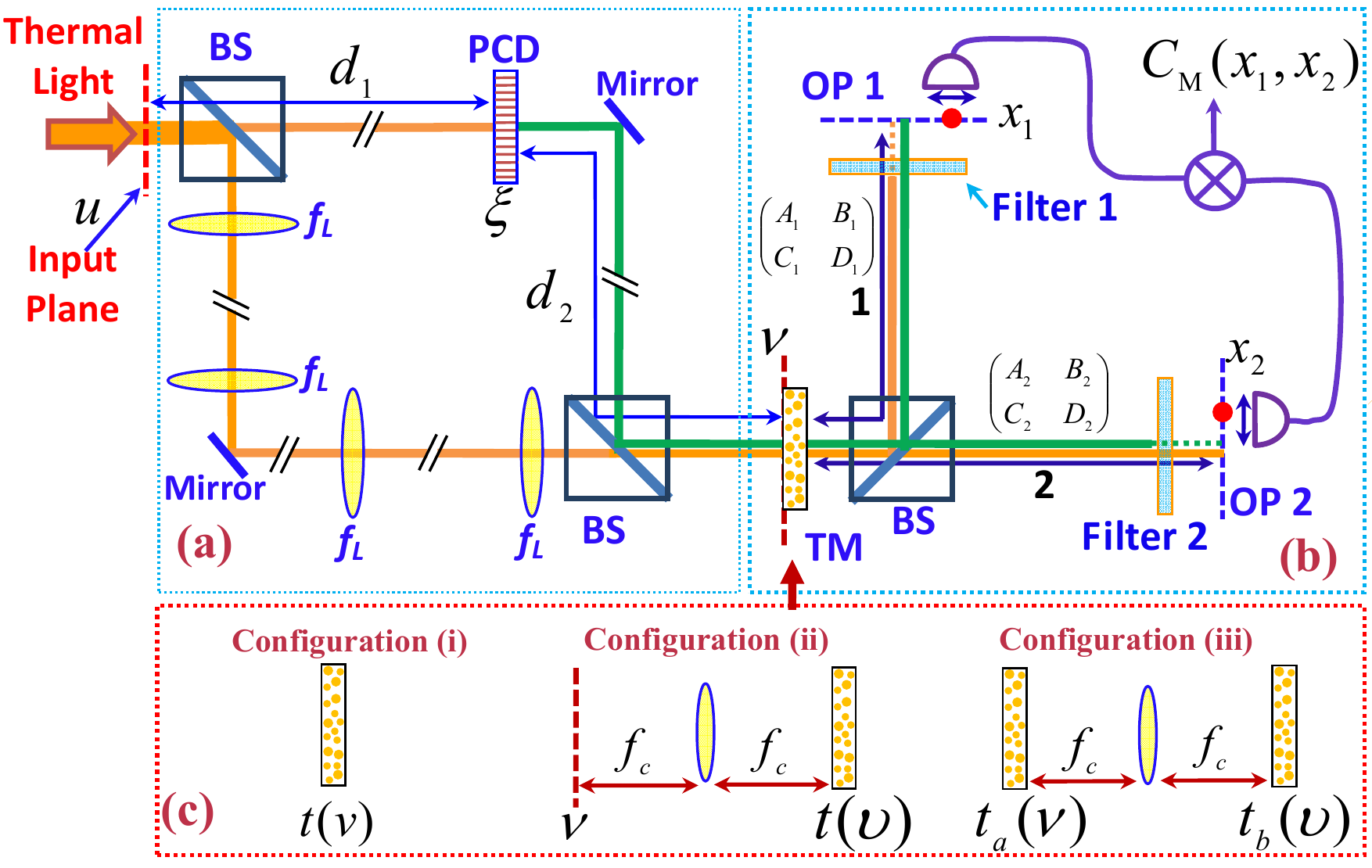}
\caption{ (Color) The modified HBT scheme for demonstrating the
intensity-correlation speckle with the phase-sensitive effects by using
thermal light fields. Part (a) provides the modified thermal source at the
incident plane ($\protect\nu $) of the TM. In the upper path of (a), the
random light fields pass through a forward non-degenerated PC device (PCD),
and the optical distances $d_{1}$($d_{2}$) between the input plane (the PCD)
and the PCD (the TM) are adjustable in order for generating the conjugated
fields at the incident plane ($\protect\nu $) of the TM. In the lower path
of (a), the random light fields pass through two pairs of 4-$f$ optical
systems, which directly image the thermal light fields onto the incident
plane of the TM. Part (b) provides the intensity-correlation measurements
similar to the case of Fig. 1, except for the filters added in each arm.
Part (c) shows three configurations (i)-(iii) for displaying three different
intensity-correlation speckles. }
\label{Fig:fig2}
\end{figure}

In order to overcome the limit of the traditional HBT-based scheme, we
design a new optical system to fulfill the phase-sensitive
intensity-intensity correlation scheme for thermal light, as shown in Fig.
2. The thermal fields first pass through the optical systems in Fig. 2(a),
for generating the modified thermal source at the incident plane ($\nu $) of
the TM [in Fig. 2(b)]. A forward non-degenerated phase conjugation (PC)
device \cite{HeGS2002} is inserted into the upper path in Fig. 2(a), and it
generates the PC waves with wavelength $\lambda _{p}$ (here $\lambda
_{p}\neq \lambda $). When $\lambda d_{1}=\lambda _{p}d_{2}$, where $d_{1}$ ($%
d_{2}$) are the distances from the input plane $u$ (the PC device) to the PC
device (the TM), then the random light at the TM via the upper path forms a
conjugated image of the input light, i. e., $E_{\nu ,\text{up}}(\nu )=\alpha
E_{i}^{\ast }(\nu )$ \cite{Note02}, where $\nu $ is the coordinate on the
incident plane of the TM, and $\alpha $ is the rate of generating the PC
light. In the lower path of Fig. 2(a), it consists of two pairs of $4-f$
optical systems \cite{Goodman1996,Pedrotti2007} with the same focus length $%
f_{L}$. Thus, the light at the TM via the lower optical path is the same as
the input field, i. e., $E_{\nu ,\text{low}}(\nu )=E_{i}(\nu )$ \cite{Note02}%
. In Fig. 2(b), it displays the measurement diagram of the
intensity-intensity correlation, and the total light fields from both two
paths of Fig. 2(a) pass through the common TM. The subsystems from the TM to
two output planes 1 and 2 also lie in Fraunhofer region (i. e., $A_{j}=0$) 
\cite{WangSMZhaoDM}, and they are the same as those in Fig.1 except for the
additional optical filters. The filters 1 and 2 transmit the light fields of
wavelength $\lambda _{p}$ and $\lambda $, respectively, while blocking the
remainder in each arm. Therefore, the intensity-intensity correlation in the
modified system can also be derived from its definition: $%
C_{M}(x_{1},x_{2})\equiv \left\langle I_{M}(x_{1})I_{M}(x_{2})\right\rangle $
\cite{MandelWolf}, where $I_{M}(x_{1,2})$ are the instantaneous intensities
on each output plane. It is the correlation between the original random
light fields and their PC fields that leads to the phase-sensitive term.
Thus, $C_{M}(x_{1},x_{2})$ now can be written as \cite{Note03}%
\begin{equation}
C_{M}(x_{1},x_{2})=\left\langle I_{M}(x_{1})\right\rangle \left\langle
I_{M}(x_{2})\right\rangle [1+\mu _{M}^{(p)}(x_{1},x_{2})],  \label{CCCMMM}
\end{equation}%
where $\mu
_{M}^{(p)}(x_{1},x_{2})=|W_{M}^{(p)}(x_{1},x_{2})|^{2}/[\left\langle
I_{M}(x_{1})\right\rangle \left\langle I_{M}(x_{2})\right\rangle ]$ is the
normalized \textit{phase-sensitive }shape function and it is dependent on
the detailed configuration of the optical system containing the TM [see Fig.
2(c)], and $W_{M}^{(p)}\!(x_{1},x_{2}\!)\!\!=\!\!\alpha \iint
\!\!W_{i}\!(\nu _{1},\nu _{2}\!)h_{1}\!(\nu _{1},x_{1}\!)h_{2}\!(\nu
_{2},x_{2}\!)d\nu _{1}\!d\nu _{2}$ is the \textit{phase-sensitive}
cross-spectral density between the two output planes in Fig. 2(b). Actually, 
$\mu _{M}^{(p)}(x_{1},x_{2})$ determines the main behavior of $%
C_{M}(x_{1},x_{2})$ since the common factor $\left\langle
I_{M}(x_{1})\right\rangle \left\langle I_{M}(x_{2})\right\rangle $ is
separable.

Next we present the results for three configurations with thermal light, as
shown in Fig. 2(c), demonstrating the similar features as two-photon speckle
patterns \cite{Peeters2010}, although the calculation is tedious but
straightforward.

In the configuration (i), the TM is located at the common imaging position
of both paths of Fig. 2(a). In this case, $\mu _{M}^{(p)}(x_{1},x_{2})$ in
Eq. (\ref{CCCMMM}) is given by \cite{Note04}%
\begin{equation}
\mu _{M}^{(p)}(x_{1},x_{2})=\frac{1}{N_{0}^{2}}\left\vert \tciFourier _{1}%
\left[ t^{2}(\nu )\right] (\frac{x_{1}}{\lambda _{p}B_{1}}+\frac{x_{2}}{%
\lambda B_{2}})\right\vert ^{2}.  \label{uuuMM}
\end{equation}%
It is clear that $\mu _{M}^{(p)}(x_{1},x_{2})$ has a different form compared
to Eq. (\ref{UUUINC}) as $|t(\nu )|^{2}$ is replaced by $t^{2}(\nu )$. The
modified intensity-intensity correlation in this case naturally contains all
phase-sensitive information of $t(\nu )$. Here the average output
intensities are $\left\langle I_{M}(x_{1})\right\rangle =I_{0}N_{0}\alpha
^{2}(\lambda _{p}\left\vert B_{1}\right\vert )^{-1}$ and $\left\langle
I_{M}(x_{2})\right\rangle =I_{0}N_{0}(\lambda \left\vert B_{2}\right\vert
)^{-1}$, which are constants and can also be subtracted from the measurement
of $C_{M}(x_{1},x_{2})$. When $\lambda _{p}B_{1}=\lambda B_{2}$, Eq. (\ref%
{uuuMM}) becomes $\mu _{M}^{(p)}(x_{1},x_{2})=N_{0}^{-2}\left\vert
\tciFourier _{1}\left[ t^{2}(\nu )\right] (\frac{x_{1}+x_{2}}{\lambda
_{p}B_{1}})\right\vert ^{2}$, i. e., a function of the sum coordinate $%
x_{1}+x_{2}$. This property is the same as that of the two-photon speckle
for the configuration (a) in Ref. \cite{Peeters2010}.

In the configuration (ii), the TM is placed at the exit plane of a 2-$f$
Fourier optical system with the focus length $f_{c}$ \cite%
{Goodman1996,Pedrotti2007}, so that $\mu _{M}^{(p)}(x_{1},x_{2})$ is given
by \cite{Note04}%
\begin{equation}
\mu _{M}^{(p)}(x_{1},x_{2})=\frac{\lambda _{p}}{\lambda N_{0}^{2}}\left\vert
\tciFourier _{1}\left[ \Omega (\upsilon )\right] (\frac{x_{2}}{\lambda B_{2}}%
-\frac{x_{1}}{\lambda B_{1}})\right\vert ^{2},  \label{uuummm22}
\end{equation}%
where $\Omega (\upsilon )=t(\upsilon )t(-\frac{\lambda _{p}}{\lambda }%
\upsilon )$ is a phase-sensitive function. From Eq. (\ref{uuummm22}), the
phase sensitive effect comes from the Fourier transformation of $\Omega
(\upsilon )$. The average intensities here are the same as that of the
configuration (i). Different from the previous case, when $B_{1}=B_{2}$, Eq.
(\ref{uuummm22}) can be rewritten as $\mu _{M}^{(p)}(x_{1},x_{2})=\frac{%
\lambda _{p}}{\lambda N_{0}^{2}}\left\vert \tciFourier _{1}\left[ \Omega
(\upsilon )\right] (\frac{x_{2}-x_{1}}{\lambda B_{1}})\right\vert ^{2}$,
which is a function of the difference coordinate $x_{2}-x_{1}$. This
property is also similar to that of the two-photon speckle for the
configuration (b) in Ref. \cite{Peeters2010}.

For the configuration (iii), two TMs are located at the incident and exit
planes of the 2-$f$ Fourier optical system with the same $f_{c}$. As pointed
out in Ref. \cite{Peeters2010}, this configuration mimics a volume
scatterer. By a tedious but straightforward calculation, $\mu
_{M}^{(p)}(x_{1},x_{2})$ is given by \cite{Note04} 
\begin{equation}
\mu _{M}^{(p)}(x_{1},x_{2})=\frac{\left\vert \tciFourier _{2}\left[ \Theta
_{p}(\upsilon _{1},\upsilon _{2})\right] (\frac{x_{1}}{\lambda _{p}B_{1}},%
\frac{x_{2}}{\lambda B_{2}})\right\vert ^{2}}{S(x_{1})S(x_{2})},
\label{uuuummm33}
\end{equation}%
where $\tciFourier _{2}$ denotes the two-dimensional Fourier transform, $%
\Theta _{p}(\upsilon _{1},\upsilon _{2})=\eta t_{b}(\upsilon
_{1})t_{b}(\upsilon _{2})\tciFourier _{1}[t_{a}^{2}(\nu )](\frac{\upsilon
_{1}}{\lambda _{p}f_{c}}+\frac{\upsilon _{2}}{\lambda f_{c}})$ with $\eta
=f_{c}^{-1}(\lambda _{p}\lambda )^{-1/2}$, and $S(x_{j})=\tciFourier
_{2}[\Theta _{n,j}(\upsilon _{1},\upsilon _{2})](-\frac{x_{j}}{\lambda
_{j}B_{j}},\frac{x_{j}}{\lambda _{j}B_{j}})$ with $\Theta _{n,j}(\upsilon
_{1},\upsilon _{2})=(\lambda _{j}f_{c})^{-1}t_{b}^{\ast }(\upsilon
_{1})t_{b}(\upsilon _{2})\tciFourier _{1}[\left\vert t_{a}(\nu )\right\vert
^{2}](\frac{\upsilon _{2}-\upsilon _{1}}{\lambda _{j}f_{c}})$. Here $%
t_{a}(\nu )$ and $t_{b}(\upsilon )$ are the complex transmission
coefficients for the two TMs, respectively; and the output average
intensities are $\left\langle I_{M}(x_{1})\right\rangle =\alpha
^{2}I_{0}(\lambda _{p}\left\vert B_{1}\right\vert )^{-1}S(x_{1})$ and $%
\left\langle I_{M}(x_{2})\right\rangle =I_{0}(\lambda \left\vert
B_{2}\right\vert )^{-1}S(x_{2})$, which are not constant any more. It is
clear that $\Theta _{p}(\nu _{1},\nu _{2})$ includes all phase information
of both $t_{a}(\nu )$ and $t_{b}(\upsilon )$, while $\Theta _{n,j}(\upsilon
_{1},\upsilon _{2})$ are phase insensitive and they are only related to the
average intensities. The difference between $\Theta _{p}(\upsilon
_{1},\upsilon _{2})$ and $\Theta _{n,j}(\upsilon _{1},\upsilon _{2})$ is the
key point for generating the phase-sensitive effect of the
intensity-intensity correlation patterns for the volume scattering phenomena
in this modified HBT scheme.

\begin{figure}[tbp]
\includegraphics[width=8.5cm]{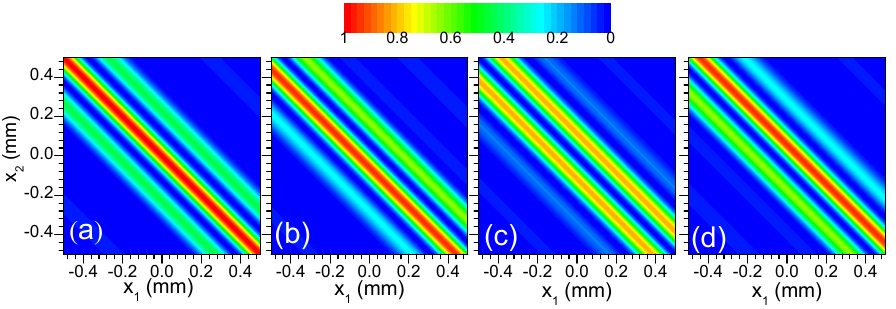}
\caption{(Color) Dependence of $\protect\mu _{M}^{(p)}(x_{1},x_{2})$ on the
phase of one of double slits. (a) $\protect\phi =0$ or $\protect\pi $, (b) $%
\protect\phi =\protect\pi /4$, (c) $\protect\phi =\protect\pi /2$, and (d) $%
\protect\phi =3\protect\pi /4$. Other parameters are $\protect\lambda %
_{p}B_{1}=\protect\lambda B_{2}=0.25$ mm$^{2}$, $a=0.5$ mm, and $b=1.0$ mm. }
\label{Fig:fig3}
\end{figure}

From the above cases, all phase information of the TMs is included in the
function $\mu _{M}^{(p)}(x_{1},x_{2})$ although different configurations may
have different specific forms. In order to understand the phase sensitive
effect in our modified scheme, we first consider a simple example--the
double slits in the configuration (i) of Fig. 2(c). The complex value of $%
t(\nu )$ for the double slits is given in Ref. \cite{Note05}. After
substituting $t(\nu )$ into Eq. (\ref{uuuMM}), we obtain $\mu
_{M}^{(p)}(x_{1},x_{2})=\sin ^{2}\left[ \phi +\pi d(\frac{x_{1}}{\lambda
_{p}B_{1}}+\frac{x_{2}}{\lambda B_{2}})\right] \sinc^{2}\left[ \pi a(\frac{%
x_{1}}{\lambda _{p}B_{1}}+\frac{x_{2}}{\lambda B_{2}})\right] $, where $a$
is the slit width, $d$ the slit separation, and $\phi $ the phase of one
slit. It is clear that the phase $\phi $ has the influence on the
distribution $\mu _{M}^{(p)}(x_{1},x_{2})$ [see Fig. 3], and different
values of $\phi $ correspond to different intensity-intensity correlation
interference patterns. From Eq. (\ref{CCCMMM}), due to the background term,
the maximal visibility of the intensity-intensity correlation interference
pattern is equal to 1/3 for the cases $\phi =m\pi $ with $m$ being an
integer. Thus, we obtain different visibility for different $\phi $. Here
only the distributions of $\mu _{M}^{(p)}(x_{1},x_{2})$ are demonstrated
since the background term $\left\langle I_{M}(x_{1})\right\rangle
\left\langle I_{M}(x_{2})\right\rangle $ can be subtracted from the
intensity correlation, like the situations in thermal ghost imaging and
interference \cite{Bennink2002,Bennink2004,Gatti2004,YanhuaShih2007}.

We now discuss the intensity-intensity correlation speckle patterns of the
thermal light passing through the different configurations in Fig. 2. Figure
4 shows the effect of the phase distribution of $t(\nu )$ on the
distribution of $\mu _{M}^{(p)}(x_{1},x_{2})$ for three different diffusers
in the configuration (i). The random amplitude and phase distributions of
three diffusers are correspondingly shown at the upper parts in Figs.
4(a)-4(c). Note that the values of $\left\vert t(\nu )\right\vert $ in Figs.
4(a)-4(c) are the same, while their phase magnitudes are totally different.
It is seen that the patterns of $\mu _{M}^{(p)}(x_{1},x_{2})$ vary with
changing the phase distributions of $t(\nu )$, and the more randomness of
the phase distributions may lead to the more homogeneous interference
speckle patterns with the smaller average speckle size.

\begin{figure}[tbp]
\includegraphics[width=8.5cm]{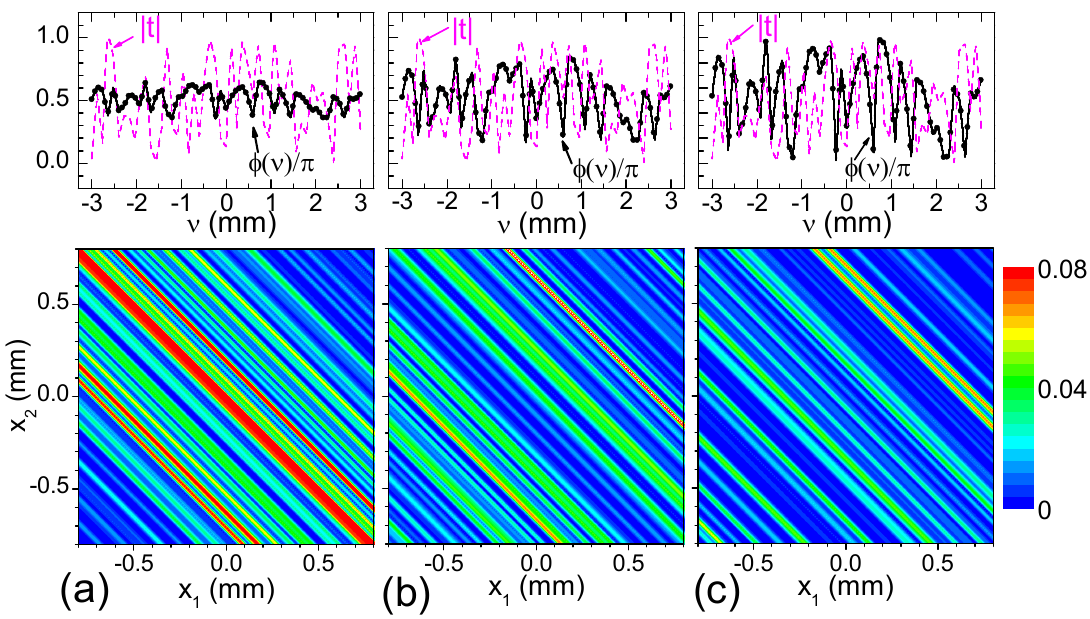}
\caption{(Color) Different distributions of $\protect\mu %
_{M}^{(p)}(x_{1},x_{2})$ for three different random diffusers in the
configuration (i) of Fig. 2(c). The corresponding upper parts show the
amplitude and phase distributions of three different\ TMs. Other parameters
are $\protect\lambda _{p}B_{1}=\protect\lambda B_{2}=0.25$ mm$^{2}$.}
\label{Fig:fig4}
\end{figure}

In Fig. 5, we demonstrate the patterns of $\mu _{M}^{(p)}(x_{1},x_{2})$ for
the diffusers in (a) the configuration (ii), and (b-c) the configuration
(iii). The functions of the TMs in these simulations are the same as that in
Fig. 4(c). Comparing with Fig. 4(c), the pattern in Fig. 5(a) is along with
the difference coordinate $x_{1}-x_{2}$ not along with the sum coordinate $%
x_{1}+x_{2}$. Such changes are similar to the cases in two-photon speckle 
\cite{Peeters2010}, and they cannot happen in the traditional HBT scheme
with thermal light. From Figs. 5(b-c), for the configuration (iii), the
patterns of $\mu _{M}^{(p)}(x_{1},x_{2})$ mimic the volume scatterer, and
the nonfactorizable features in the correlation patterns are clearly seen.
For a small value of $f_{c}$ in Fig. 5(c), the correlation speckle spots in
the pattern of $\mu _{M}^{(p)}(x_{1},x_{2})$ are elongated along the
difference coordinate of $x_{1}-x_{2}$. This can be understood from the fact
that the second diffuser is illuminated with the far-field patterns of the
first diffuser. Within the same area of the second diffuser, the smaller of $%
f_{c}$, the less information from the first diffuser can be projected. This
can be seen from the form of the function $\Theta _{p}(\nu _{1},\nu _{2})$.
Therefore, we can conclude that the modified HBT scheme with thermal light
can provide the \textit{phase-sensitive} intensity-intensity correlation
speckle.\ 

\begin{figure}[tbp]
\includegraphics[width=8.5cm]{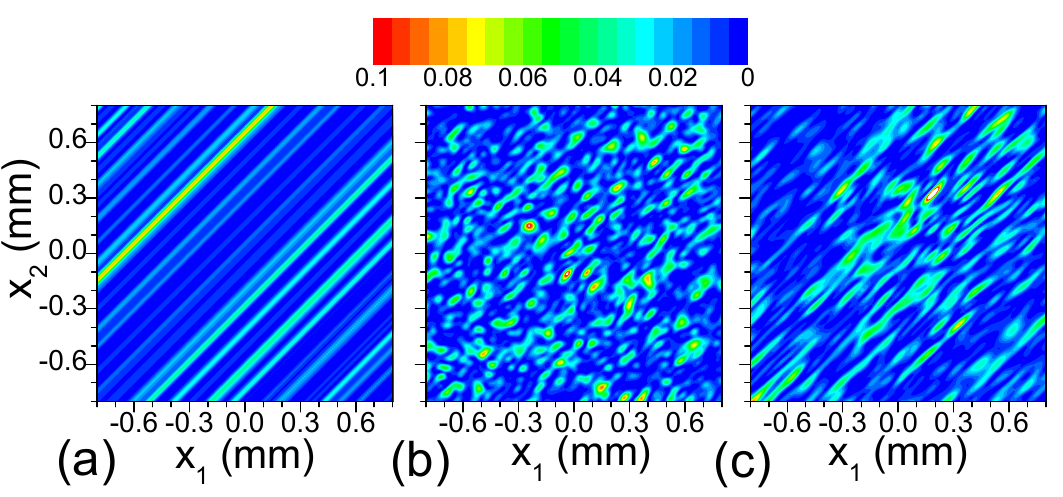}
\caption{(Color) Distributions of $\protect\mu _{M}^{(p)}(x_{1},x_{2})$ for
the random diffusers in the cases of (a) the configuration (ii), and (b-c)
the configuration (iii). Other parameters are $\protect\lambda _{p}=550$nm, $%
\protect\lambda =500$nm, $B_{1}=B_{2}=500$ mm, $f_{c}=500$mm for (b) and $%
150 $mm for (c).}
\label{Fig:fig5}
\end{figure}

Lastly, we discuss the possibility of experimentally realizing our scheme.
The key challenge of our scheme in Fig. 2 is to generate the non-degenerate
PC fields of thermal light. For demonstrating our predicted result, one can
employ the psudothermal light source (produced via the random scattering
when a laser field passes through a ground glass) as the input light. The PC
light of the psudothermal light can be generated via the conventional PC
technologies, such as the four-wave mixing processes (e. g., Refs. \cite%
{Hellwarth1977, Bloom1977,Heer1979,Khyzniak1984}) and the stimulated
scattering processes (e. g., Refs. \cite{Yu1972,Wang1978,
Mullen1992,Bowers1998}). For example, the nondegenerate PC light is
generated by using a Pr$^{3+}$:Y$_{2}$SiO$_{5}$ crystal based on the
electromagnetically induced transparency effect \cite{ZhaiZH2011}.
Meanwhile, the fidelity of the PC fields may have an influence on the
correlations between the input and PC fields, and this will in turn affect
the intensity-intensity correlations. In another scheme, we can use the
novel digital PC technology \cite{Cui2010, Hsieh2010, Wang2012, Hillman2013}%
, which does not involve the nonlinear processes and can even generate the
high-quality PC waves for the weak, incoherent fluorescence signal \cite%
{Vellekoop2013}, to verify this effect. In fact, if the filters in Fig. 2(b)
are removed or disabled (when $\lambda _{p}=\lambda $), both the
phase-sensitive and phase-insensitive terms will occur in Eq. (\ref{CCCMMM}%
), which only increases the complexity to determine the phase-sensitive
patterns.

In summary, we have presented the phase-sensitive intensity-intensity
correlation speckle effect of thermal light in the modified HBT scheme. This
scheme is based on introducing the PC light to change the correlations
between the two optical paths. It is revealed that the phase-sensitive and
nonfactorizable features can be seen in thermal intensity-intensity
correlation speckle. Finally, the discussion on the experimental realization
is presented. This scheme is different from those thermal ghost imaging and
diffraction \cite%
{Bennink2002,Bennink2004,Gatti2004,YanhuaShih2007,Borghi2006} and the
unbalanced interferometer-based scheme via the direct intensity measurements 
\cite{Zhang2009}, since all thermal photons in our case pass through the
common sample. Our scheme can also be used to recover the phase information
in the thermal-like temporal intensity-intensity correlation cases \cite%
{Torres-Company2012}. This modified HBT scheme may have important
applications for developing the intensity-intensity correlation speckle and
imaging technologies of thermal or incoherent light sources.

\begin{acknowledgments}
This work is supported by NPRP grant 4-346-1-061 by the Qatar National
Research Fund and a grant from King Abdulaziz City for Science and
Technology. This research is also supported by NSFC grants (No. 11274275 and
No. 61078021), and the grant by the National Basic Research Program of China
(No. 2012CB921602).
\end{acknowledgments}

\end{document}